\documentclass[final,3p,times,twocolumn]{elsarticle}
 \biboptions{comma,sort&compress}
\usepackage{ecrc}
\usepackage{here}
 \usepackage{graphicx}
 \usepackage{epsfig}
\usepackage{epstopdf}
\usepackage{amsmath}

\def\nin{\noindent}
\def\beq{\begin{equation}}
\def\eeq{\end{equation}}
\def\bea{\begin{eqnarray}}
\def\eea{\end{eqnarray}}
\def\nnb{\nonumber}
\def\la{\langle}
\def\ra{\rangle}
\def\ga{\left(}
\def\dr{\right)}

\usepackage{graphicx}
\usepackage{here}
\def\beq{\begin{equation}}
\def\eeq{\end{equation}}
\def\bea{\begin{eqnarray}}
\def\eea{\end{eqnarray}}
\def\bq{\begin{quote}}
\def\eq{\end{quote}}

\def\nnb{\nonumber}
\def\ga{\left(}
\def\dr{\right)}

\def\nnb{\nonumber}
\def\la{\langle}
\def\ra{\rangle}
\def\nin{\noindent}
\def\ba{\begin{array}}
\def\ea{\end{array}}

\def\b{$\bullet~$}
\def\als{\alpha_s}

\def\gg2{ \la\alpha_s G^2 \ra}
\def\gg3{g^3f_{abc}\la G^aG^bG^c \ra}
\def\ggg4{\la\als^2G^4\ra}

\def\beq{\begin{equation}}
\def\enq{\end{equation}}
\def\beqa{\begin{eqnarray}}
\def\enqa{\end{eqnarray}}
\def\nnb{\nonumber}

\def\qq{\lag\bar{q}q\rag}

\def\lb{\label}

\newcommand{\rag}{\rangle}
\newcommand{\lag}{\langle}

\def\ln{\mbox{Log}}
\def\gg{\lag g^{2}_{s} G^2 \rag}
\def\ggg{\lag g^{3}_{s}G^3\rag}

\volume{00}
\firstpage{1}
\journalname{Nuclear and Particle Physics Proceedings }
\runauth{Albuquerque et al.}
\jnltitlelogo{Nuclear and Particle Physics Proceedings }

\begin{document}

\begin{frontmatter}



\title{ XYZ - SU3 Breakings from Laplace Sum Rule at Higher Orders: Summary \tnoteref{text1}}


\author{R. Albuquerque}
\address{Faculty of Technology,Rio de Janeiro State University (FAT,UERJ), Brazil}
\ead{raphael.albuquerque@uerj.br}
\author{S. Narison 
}
\address{Laboratoire
Univers et Particules de Montpellier (LUPM), CNRS-IN2P3, \\
Case 070, Place Eug\`ene
Bataillon, 34095 - Montpellier, France.}
\ead{snarison@yahoo.fr}

\author{
D. Rabetiarivony
\fnref{label1}}\ead{rd.bidds@gmail.com}
\author{ G. Randriamanatrika\fnref{label1}}\ead{artesgaetan@gmail.com}

\address{Institute of High-Energy Physics of Madagascar (iHEPMAD)\\
University of Ankatso,
Antananarivo 101, Madagascar}

\fntext[label1]{Ph.D student}
\tnotetext[text1]{Part of a review presented by S. Narison @  QCD17 (3-7july 2017, Montpellier-FR) and Talks given by D. Rabetiarivony and G. Randriamanatrika @ HEPMAD17 (21 - 26 September, Antananarivo - MG)}





\begin{abstract}
\noindent
This talk reviews and summarizes some of our results in \cite{SU3} on XYZ- SU3 Breakings obtained from QCD Laplace Sum Rules (LSR) at next-to-next-leading order (N2LO) of perturbative (PT) theory and including next-to-leading order (NLO) SU3 breaking corrections and leading order (LO) contributions of dimensions $d\leq (6 - 8)$ non-perturbative condensates. We conclude that the observed X states are good candidates for being $1^{++}$ and $0^{++}$ molecules states. We find that the SU3 breakings are relatively small for the masses ($\leq 10$ (resp. 3)  $\%$) for the charm (resp. bottom) channels while they are large ($\leq 20\, \%$) for the couplings. Like in the chiral limit case, the couplings decrease faster: $1/m_{b}^{3/2}$ than $1/m_{b}^{1/2}$ of HQET. Our approach cannot clearly separate ( within the errors ) some molecule states from the four-quark ones with the same quantum numbers.
\end{abstract}

\scriptsize
\begin{keyword}
QCD Spectral Sum Rules, Perturbative and Non-perturbative QCD, Exotic hadrons, Masses and Decay constants.


\end{keyword}

\end{frontmatter}


\section{Introduction}
In recent papers \cite{SU3,SNX2, SNX1,SNCHI2,X3D}, we have used QCD spectral ( Laplace \cite{SVZa, BELL,BERT,NEUF} and FESR \cite{FESRa}) sum rules \cite{SNB1, SNB2,CK} to improve some previous LO results for the masses and decay constants of the XYZ exotic heavy-light and charmonium-like mesons obtained in the chiral limit \cite{X3D,X1A, X3B,NIELSEN}. In so doing, we include the SU3 NLO PT corrections into the N2LO PT factorizable  chiral limit corrections to the heavy-light exotic correlators. To these higher order (HO) PT contributions, we add the LO contribution of condensates having a dimension ($d\leq 6$). Like in the chiral limit case\,\cite{SNX2}, we do not include into the analysis contributions of condensates of higher dimension ($d\geq 8$) but only consider their effects as a source of systematic errors due to the truncation of the Operator Product Expansion (OPE). 

Recent measurements of the $J/\psi \phi$ invariant masses from $B^{+}\rightarrow$ $ J/\psi \phi K^{+}$ decays by the LHCb collaboration\,\cite{LHCb1} confirmed the existence of the $X(4147)$ and $X(4273)$ with quantum numbers $1^{++}$ found earlier by the CDF\,\cite{CDFX, CDFX1}, the CMS\,\cite{CMS} and the D0\,\cite{D0} collaborations. In the same time, the LHCb collaboration has reported the existence of the $0^{++}$ states in the analogous $J/\psi \phi$ invariant masses.
\section{Molecules and four-quark two point functions}
We shall work with the transverse part $\Pi^{(1)}_{mol}$ of the two-point spectral functions\,\footnote{Hereafter, similar expressions will be obtained for the four-quark states by replacing the sub-index {\it mol} by {\it 4q}.}:
\bea
\hspace*{-0.35cm}\Pi^{\mu\nu}_{mol}(q)\hspace{-0.3cm}&\equiv&\hspace{-0.3cm} i\int d^4x ~e^{iq.x}\lag 0
|T[{\cal O}_{mol}^{\mu}(x){\cal O}_{mol}^{\nu\dagger}(0)]
|0\rag\nnb\\
&=&\hspace*{-0.35cm} -\Pi^{(1)}_{mol}(q^2) \left(g^{\mu\nu}-\frac{q^\mu q^\nu}{q^2} \right)+\Pi^{(0)}_{mol}(q^2)\frac{q^\mu
q^\nu}{ q^2},
\lb{eq:2pf}
\eea
for the spin $1$ states while for the spin $0$ ones, we shall use the two-point functions $\psi_{mol}(q^2)$ built directly from the (pseudo)scalar currents:
\beq 
\psi_{mol}(q^2)=i\int d^4x ~e^{iq.x}\lag 0
|T[{\cal O}_{mol}(x){\cal O}_{mol}(0)]
|0\rag,
\label{eq:s02pf}
\eeq
which is related to $\Pi^{(0)}$ appearing in Eq.(\ref{eq:2pf}) via Ward identities \cite{SNB1, SNB2,BECCHI}.
Thanks to their analyticity properties $\Pi^{(1,0)}_{mol}$ and $\psi_{mol}$ obey the dispersion relation:
\begin{align}
\Pi^{(1,0)}_{mol}(q^2) ;\psi_{mol}(q^2) &= \frac{1}{\pi}\int_{4 M^2_Q}^{\infty} \hspace*{-0.25cm}dt ~\frac{{\rm Im}\Pi^{(1,0)}_{mol}(t);{\rm Im}\psi_{mol}(t)}{t-q^2-i \epsilon},
\end{align}
where Im$\Pi^{(1,0)}_{mol}(t)$ and Im$\psi_{mol}(t)$ are the spectral functions.
\subsection*{$\bullet$ Interpolating currents}
\nin
The interpolating current ${\cal O}_{mol}$ for the molecules  and ${\cal O}_{4q}$ for four-quark states  are given in Table \ref{tab:curmol} and Table \ref{tab:cur4q}.
 {\scriptsize
\begin{table}[hbt]
\setlength{\tabcolsep}{0.7pc}
 \caption{\footnotesize    
Interpolating currents ${\cal O}_{mol}$ with a definite C-parity describing the molecule-like states. $Q\equiv c$(resp. b) for the $\bar{D}_s D_s$  (resp. $\bar{B}_s B_s$)-like molecules.}
 {\scriptsize
\begin{tabular}{lll}
\hline
\hline
\\
States&$J^{PC}$&Molecule currents$\equiv {\cal O} _{mol}(x)$ \\
\\
\hline
{\bf Scalar}&$0^{++}$&\\
$\bar{D}_s D_s\, ,\, \bar{B}_sB_s$&&$(\bar{s}\gamma_5 Q)(\bar{Q}\gamma_5 s)$\\
$\bar{D}^*_s D^*_s\, ,\, \bar{B}^*_sB^*_s$&&$(\bar{s}\gamma_{\mu} Q)(\bar{Q}\gamma^{\mu} s)$\\
$\bar{D}^*_{s0} D^*_{s0}\, ,\, \bar{B}^*_{s0}B^*_{s0}$&&$(\bar{s}Q)(\bar{Q}s)$\\
$\bar{D}_{s1} D_{s1}\, ,\, \bar{B}_{s1}B_{s1}$&&$(\bar{s}\gamma_{\mu} \gamma_5 Q)(\bar{Q}\gamma^{\mu} \gamma_5 s)$\\
{\bf Axial-vector}&$1^{++}$&\\
$\bar{D}^*_s D_s\, ,\, \bar{B}^*_sB_s$&&$\frac{i}{\sqrt{2}}\left[(\bar{Q}\gamma_{\mu}s)(\bar{s}\gamma_5 Q)-(\bar{s}\gamma_{\mu} Q)(\bar{Q}\gamma_5 s)\right]$\\
$\bar{D}^*_{s0} D_{s1}\, ,\, \bar{B}^*_{s0}B_{s1}$&&$\frac{i}{\sqrt{2}}\left[(\bar{s}Q)(\bar{Q}\gamma_{\mu}\gamma_5 s)+(\bar{Q}s)(\bar{s}\gamma_{\mu}\gamma_5 Q)\right]$\\
\bf Pseudoscalar&$0^{- \pm}$& \\
$D^*_{s0}D_s$,$B^*_{s0}B_s$&&$ \frac{1}{ \sqrt{2}}  \bigg[
		\big(  \bar{s}Q \big) \big( \bar{Q} \gamma_5 s \big) 
		\pm \:\big(  \bar{Q}s \big) \big( \bar{s} \gamma_5 Q \big)  \bigg]$\\
$D^*_{s}D_{s1}$,$B^*_{s}B_{s1}$&&$ \frac{1}{ \sqrt{2}}  \bigg[
		\big(  \bar{Q}\gamma_{\mu}s \big) \big( \bar{s} \gamma^{\mu} \gamma_5 Q \big) 
		\mp \:\big(  \bar{Q}\gamma_{\mu} \gamma_5 s \big) \big( \bar{s} \gamma^{\mu} Q \big)  \bigg]$  \\
\bf vector&$1^{- \pm}$&\\
$D^*_{s0}D^*_s$,$B^*_{s0}B^*_s$&&$ \frac{1}{ \sqrt{2}}  \bigg[
		\big(  \bar{s}Q \big) \big( \bar{Q} \gamma_{\mu} s \big) 
		\mp \:\big(  \bar{Q}s \big) \big( \bar{s} \gamma_{\mu} Q \big)  \bigg] $  \\
$D_{s}D_{s1}$,$B_{s}B_{s1}$&&$\frac{1}{ \sqrt{2}}  \bigg[
		\big(  \bar{Q} \gamma_{\mu} \gamma_5 s \big) \big( \bar{s}  \gamma_5 Q \big) 
		\pm \:\big(  \bar{s}\gamma_{\mu} \gamma_5 Q \big) \big( \bar{Q} \gamma_5 s \big)  \bigg] $   \\
\hline\hline
\end{tabular}
}
\label{tab:curmol}
\end{table}
}
 {\scriptsize
\begin{table}[hbt]
\begin{center}
\setlength{\tabcolsep}{.2pc}
 \caption{\footnotesize    
Interpolating currents describing the four-quark states. $Q\equiv c$ (resp $b$). $k$ is an arbitrary current mixing where the optimal value is found to be $k=0$ from \cite{X3B}}
 {\scriptsize
\begin{tabular}{lll}
\hline
\hline
\\
States&$J^{P}$&Four-quark currents$\equiv {\cal O} _{4q}(x)$ \\
\\
\hline
\\
{\bf Scalar}&$0^{+}$&$\epsilon_{abc}\epsilon_{dec}\left[ (s^{T}_{a}C\gamma_5 Q_b)(\bar{s}_d \gamma_5 C \bar{Q}^{T}_{e})+k(s^{T}_{a} C Q_b)(\bar{s}_d  C \bar{Q}^{T}_{e})\right]$\\
{\bf Axial-vector}&$1^{+}$&$\epsilon_{abc}\epsilon_{dec}\left[ (s^{T}_{a}C\gamma_5 Q_b)(\bar{s}_d \gamma_{\mu} C \bar{Q}^{T}_{e})+k(s^{T}_{a} C Q_b)(\bar{s}_d  \gamma_{\mu} \gamma_5 C \bar{Q}^{T}_{e})\right]$\\
{\bf Pseudoscalar}&$0^{-}$&$\epsilon_{abc}\epsilon_{dec}\left[ (s^{T}_{a}C\gamma_5 Q_b)(\bar{s}_d C \bar{Q}^{T}_{e})+k(s^{T}_{a} C Q_b)(\bar{s}_d  \gamma_5 C \bar{Q}^{T}_{e})\right]$\\
{\bf Vector}&$1^{-}$&$\epsilon_{abc}\epsilon_{dec}\left[ (s^{T}_{a}C\gamma_5 Q_b)(\bar{s}_d \gamma_{\mu} \gamma_5 C \bar{Q}^{T}_{e})+k(s^{T}_{a} C Q_b)(\bar{s}_d  \gamma_{\mu} C \bar{Q}^{T}_{e})\right]$\\
\hline\hline
\end{tabular}
}
\label{tab:cur4q}
\end{center}
\vspace*{-0.5cm}
\end{table}
}
\subsection*{$\bullet$ Spectral function within MDA}
\nin
We shall use the Minimal Duality Ansatz (MDA) given in Eq.\ref{eq:mda} for parametrizing the spectral function:
\beq
\frac{1}{\pi}\hspace{-0.1cm}\mbox{ Im}\Pi_{mol}(\hspace{-0.04cm} t \hspace{-0.03cm}) \hspace{-0.1cm}\simeq \hspace{-0.15cm}f^2_{H}M^8_{H}\delta(\hspace{-0.05cm}t-M_{H}^2)+\mbox{``QCD continuum"}\theta (\hspace{-0.04cm}t-t_c\hspace{-0.05cm}),
\label{eq:mda}
\eeq
where $f_H$ is the decay constant defined as:
\beq
\la 0| {\cal O}_{mol}|H\ra\hspace{-0.08cm}=\hspace{-0.08cm}f_{H}M^4_{H}~,~\la 0| {\cal O}_{mol}^\mu|H\ra\hspace{-0.08cm}=\hspace{-0.08cm}f_{H}M^5_{H}\epsilon_\mu,
\label{eq:coupling}
\eeq
respectively for spin 0 and 1 states with  $\epsilon_\mu$ the (axial-)vector polarization. The higher order states contributions are smeared by the ``QCD continuum" coming from the discontinuity of the QCD diagrams and starting from a constant threshold $t_c$.
\subsection*{$\bullet$ NLO and N2LO PT corrections using factorization}
Assuming a factorization of the four-quark interpolating current as a natural consequence of the molecule definition of the state, we can write the corresponding spectral function as a convolution of the spectral functions associated to quark bilinear current. In this way, we obtain\cite{PICH} for the $\bar{D}D^*$ and $\bar{D}^*_0 D^*$-like spin 1 states:
\bea
\frac{1}{ \pi}{\rm Im} \Pi^{(1)}_{mol}(t)\hspace{-0.3cm}&=&\hspace{-0.3cm}\theta (t-4M_Q^2) \hspace{-0.05cm} \ga \hspace{-0.05cm}\frac{1}{ 4\pi}\hspace{-0.05cm}\dr^2 \hspace{-0.12cm} t^2\hspace{-0.2cm} \int_{M_Q^2}^{(\sqrt{t}-M_Q)^2} \hspace*{-1cm}dt_1 \hspace{0.25cm}\int_{M_Q^2}^{(\sqrt{t}-\sqrt{t_1})^2} \hspace*{-1cm}dt_2\nnb\\
&&\hspace{-0.3cm}\times\lambda^{3/2}\frac{1}{ \pi}{\rm Im} \Pi^{(1)}(t_1) \frac{1}{ \pi}{\rm Im} \psi^{(s,p)}(t_2)~.
\label{eq:convolution}
\eea
For the $\bar{D}D$ spin $0$ state, one has:
\bea
\frac{1}{ \pi}{\rm Im} \psi^{(s)}_{mol}(t)\hspace{-0.08cm}=\hspace{-0.08cm}\theta (t-4M_Q^2) \hspace{-0.05cm} \ga \hspace{-0.05cm}\frac{1}{ 4\pi}\hspace{-0.05cm}\dr^2 \hspace{-0.12cm} t^2\hspace{-0.2cm} \int_{M_Q^2}^{(\sqrt{t}-M_Q)^2} \hspace*{-1cm}dt_1 \hspace{0.25cm}\int_{M_Q^2}^{(\sqrt{t}-\sqrt{t_1})^2} \hspace*{-1.1cm}dt_2 \nnb\\
\times\lambda^{1/2} \ga  \frac{t_1}{ t}\hspace{-0.1cm}+\frac{t_2}{ t}-1  \dr^2\frac{1}{ \pi}{\rm Im}\psi^{(p)}(t_1) \frac{1}{ \pi} {\rm Im} \psi^{(p)}(t_2),
\eea

and for the $\bar{D}^*D^*$ spin $0$ state:
\bea
\frac{1}{ \pi}{\rm Im} \psi^{(s)}_{mol}(t)\hspace{-0.08cm}=\hspace{-0.08cm}\theta (t-4M_Q^2) \hspace{-0.05cm} \ga \hspace{-0.05cm}\frac{1}{ 4\pi}\hspace{-0.05cm}\dr^2 \hspace{-0.12cm} t^2\hspace{-0.2cm} \int_{M_Q^2}^{(\sqrt{t}-M_Q)^2} \hspace*{-1cm}dt_1 \hspace{0.25cm}\int_{M_Q^2}^{(\sqrt{t}-\sqrt{t_1})^2} \hspace*{-1cm}dt_2\nnb\\\times\lambda^{1/2}\times\left[\ga\frac{t_1}{ t}+\frac{t_2}{ t}-1\dr^2+\frac{8t_1 t_2}{t^2}\right]\nnb\\
\times\frac{1}{ \pi}{\rm Im}\Pi^{(1)}(t_1) \frac{1}{ \pi} {\rm Im} \Pi^{(1)}(t_2),
\eea

where:
\beq
\lambda=\ga 1-\frac{\ga \sqrt{t_1}- \sqrt{t_2}\dr^2}{ t}\dr \ga 1-\frac{\ga \sqrt{t_1}+ \sqrt{t_2}\dr^2}{ t}\dr~,
\eeq
is the phase space factor and $M_Q$ is the on-shell heavy quark mass. ${\rm Im}\Pi^{(1)}(t) $ is the spectral function associated to the bilinear $\bar{c}\gamma_{\mu}(\gamma_5)q$ (axial-)vector current, while ${\rm Im}\,\psi^{(s,p)}(t) $ is associated to the $\bar{c}i(\gamma_5)q$ (pseudo)scalar current\footnote{In the chiral limit $m_q=0$, the PT expressions of the vector (resp. scalar) and axial-vector (resp. pseudoscalar) spectral functions are the same.}. We shall assume that a suchfactorization also holds for four-quark states. 
\subsection*{$\bullet$ The inverse Laplace transform sum rule (LSR)}
The LSR and its ratio read:
\beq
{\cal L}_{mol}(\tau,t_c,\mu)=\frac{1}{\pi}\!\int_{4M_Q^2}^{t_c}\!dt\,e^{-t\tau}\mbox{Im}\{\Pi_{mol} ; \psi_{mol}\}(t,\mu),
\label{eq:LSR}
\eeq
\beq
{\cal R}_{mol}(\tau,t_c,\mu)\! =\! \frac{\int_{4M_Q^2}^{t_c}\! dt\,t\,e^{-t\tau}\mbox{Im}\{\Pi_{mol} ; \psi_{mol}\}(t,\mu)}
{\int_{4M_Q^2}^{t_c}\! dt\, e^{-t\tau} \mbox{Im}\{\Pi_{mol} ; \psi_{mol}\}(t,\mu)}\!\simeq \! M_R^2,
\eeq
where $\mu$ is the subtraction point which appears in the approximate QCD series when radiative corrections are included and $\tau$ is the sum rule variable replacing $q^2$.
\subsection*{$\bullet$ Double ratios of inverse Laplace transform sum rule}
Double Ratios of Sum Rules (DRSR)\,\cite{DRSR,SNFORM1,SNGh3,
HBARYON1,HBARYON2,NAVARRA,SNB1,SNB2} 
are also useful for extracting the SU3 breaking effects on couplings and mass ratios. They read:
\beq
f^{sd}_{mol}\equiv \frac{{\cal L}_{mol}^{s}(\tau,t_c,\mu)}{{\cal L}_{mol}^{d}(\tau,t_c,\mu)}~, ~~~r_{mol}^{sd}\equiv \frac{{\cal R}_{mol}^{s}(\tau,t_c,\mu)}{{\cal R}_{mol}^{d}(\tau,t_c,\mu)},
\label{eq:DRSR}
\eeq
the upper indices $s,d$ indicates the $s$ and $d$ quark channels. These DRSR can be used when each sum rule optimizes at the same values of the parameters $(\tau,t_c,\mu)$.

\subsection*{$\bullet$ Stability criteria and some phenomenological tests}
The variables $\tau,\mu$ and $t_c$ are, in principle, free external parameters. We shall use stability criteria (if any) with respect to these free 3 parameters, for extracting the optimal results. In the standard MDA given in Eq.\ref{eq:mda} for parametrizing the spectral function, the ``QCD continuum" threshold $t_c$ is constant and is independent on the subtraction point $\mu$. One should notice that this standard MDA with constant $t_c$ describes quite well the properties of the lowest ground state as explicitly demonstrate in \cite{SNFB12a} and in various examples \cite{SNB1, SNB2} after confronting the integrated spectral function within this simple parametrization with the full data measurments. It has been also successfully tested in the large $N_c$ limit of QCD in \cite{PERISb}. Though it is difficult to estimate with a good precision the systematic error related to this simple model for reproducing accurately the data, we expect that the same feature is reproduced for the case of the XYZ discussed here where complete data are still lacking.
\section{QCD input parameters}
The QCD parameters which shall appear in the following analysis will be the charm and bottom quark masses $m_{c,b}$, the strange quark mass $m_s$ (we shall neglect  the light quark masses $m_{u,d}$), the light quark condensate $\qq$ ($q\equiv u,d$),  the gluon condensates $ \lag\alpha_sG^2\rag \equiv \la \alpha_s G^a_{\mu\nu}G_a^{\mu\nu}\ra$ 
and $ \la g^3G^3\ra \equiv \la g^3f_{abc}G^a_{\mu\nu}G^b_{\nu\rho}G^c_{\rho\mu}\ra$, 
the mixed condensate $\la\bar qGq\ra \equiv {\la\bar qg\sigma^{\mu\nu} (\lambda_a/2) G^a_{\mu\nu}q\ra}=M_0^2\la \bar qq\ra$ 
and the four-quark  condensate $\rho\alpha_s\la\bar qq\ra^2$, where  $\rho\simeq 3-4$ indicates the deviation from the four-quark vacuum saturation. Their values are given in Table \ref{tab:param} and more recently confirmed in\,\cite{SN18}. The original errors on 
$\kappa\equiv \la \bar ss\ra/\la\bar dd\ra$ have been enlarged to take into account the lattice result\,\cite{MCNEILE} which needs to be checked by some other groups. 
We shall work with the running light quark condensates, which read to leading order in $\alpha_s$: 
 \bea
{\la\bar qq\ra}(\tau)&=&-{\hat \mu_q^3  \ga-\beta_1a_s\dr^{2/{\beta_1}}},\nnb\\
{\la\bar q Gq\ra}(\tau)&=&-{M_0^2{\hat \mu_q^3} \ga-\beta_1a_s\dr^{1/{3\beta_1}}},
\label{d4g}
\eea
and the running quark mass to NLO (for the number of flavours $n_f=3$)
\beq
\overline{m}_s(\tau)=\hat m_s(-\beta_1 a_s)^{-2/\beta_1}(1+0.8951a_s),
\eeq
where $\beta_1=-(1/2)(11-2n_f/3)$ is the first coefficient of the $\beta$ function 
for $n_f$ flavours; $a_s\equiv \alpha_s(\tau)/\pi$; $\hat\mu_q$ and $\hat m_s$ is the spontaneous RGI light quark condensate \cite{FNR} and strange quark mass.

{\scriptsize
\begin{table}[h]
\setlength{\tabcolsep}{.2pc}
 \caption{QCD input parameters:the original errors for $\la\alpha_s G^2\ra$, $\la g^3  G^3\ra$ and $\rho \la \bar qq\ra^2$ have been multiplied by about a factor 3 for a conservative estimate of the errors (see also the text). }  
    {\footnotesize
 {\begin{tabular}{@{}lll@{}}
&\\
\hline
\hline
Parameters&Values& Ref.    \\
\hline
$\alpha_s(M_\tau)$& $0.325(8)$&\cite{SNTAU,BNPa,BNPb,BETHKE}\\
$\hat m_s$&$(0.114\pm0.006)$ GeV &\cite{SNB1,SNTAU,SNmassa,SNmassb,SNmass98a,SNmass98b,SNLIGHT}\\
$\overline{m}_c(m_c)$&$1261(12)$ MeV &average \cite{SNmass02,SNH10a,SNH10b,SNH10c,PDG,IOFFEa,IOFFEb}\\
$\overline{m}_b(m_b)$&$4177(11)$ MeV&average \cite{SNmass02,SNH10a,SNH10b,SNH10c,PDG}\\
$\hat \mu_q$&$(253\pm 6)$ MeV&\cite{SNB1,SNmassa,SNmassb,SNmass98a,SNmass98b,SNLIGHT}\\
$\kappa\equiv \la \bar ss\ra/\la\bar dd\ra$& $(0.74^{+0.34}_{- 0.12})$&\cite{HBARYON1,HBARYON2,SNB1}\\
$M_0^2$&$(0.8 \pm 0.2)$ GeV$^2$&\cite{JAMI2a,JAMI2b,JAMI2c,HEIDb,HEIDc,SNhl}\\
$\la\alpha_s G^2\ra$& $(7\pm 3)\times 10^{-2}$ GeV$^4$&
\cite{SNTAU,LNT,SNIa,SNIb,YNDU,SNH10a,SNH10b,SNH10c,SNG2,SNGH}\\
$\la g^3  G^3\ra$& $(8.2\pm 2.0)$ GeV$^2\times\la\alpha_s G^2\ra$&
\cite{SNH10a,SNH10b,SNH10c}\\
$\rho \alpha_s\la \bar qq\ra^2$&$(5.8\pm 1.8)\times 10^{-4}$ GeV$^6$&\cite{SNTAU,LNT,JAMI2a,JAMI2b,JAMI2c}\\
\hline\hline
\end{tabular}}
}
\label{tab:param}
\vspace*{-0.5cm}
\end{table}
} 
\section{QCD expressions of the spectral functions}
In our works \cite{SNX1,SNX2,SU3}, we provide new compact integrated expressions of QCD spectral functions at LO of PT QCD and including non-perturbative (NP) condensates having dimensions $d\leq 6-8$. NLO and N2LO corrections are introduced using the convolution integrals in Eq.\,\ref{eq:convolution}. The expressions of QCD spectral functions of heavy-light bilinear currents are known to order $\alpha_s$ (NLO) from\, \cite{BROAD} and to order $\alpha^2_s$ (N2LO) in the chiral limit $m_q=0$ from\,\cite{CHETa,CHETb} which are available as a Mathematica program named Rvs. We shall use the SU3 breaking PT corrections at NLO\, \cite {GELH} from the two-point function formed by bilinear currents. N3LO corrections are estimated from a geometric growth of the QCD PT series\,\cite{SNZ} as a source of PT errors, which we expect to give a good approximation of the uncalculated higher order terms dual to the $1/q^2$ contribution of a tachyonic gluon mass\,\cite{CNZ1, CNZ2}.

In our analysis, we replace the on-shell (pole) mass appearing in the spectral functions with the running mass using the relation, to order $\alpha^2_s$\,\cite{SNB1,SNB2,SNB3}:
\bea
M_Q \hspace{-0.3cm}&=&\hspace{-0.3cm} \overline{m}_Q(\mu)\Big{[}1+\frac{4}{3} a_s+ (16.2163 -1.0414 n_l)a_s^2\nnb\\
&&\hspace{-0.3cm}+\ln{\ga\frac{\mu}{ M_Q}\dr^2} \ga a_s+(8.8472 -0.3611 n_l) a_s^2\dr\nnb\\
&&\hspace{-0.3cm}+\ln^2{\ga\frac{\mu}{ M_Q}\dr^2} \ga 1.7917 -0.0833 n_l\dr a_s^2...\Big{]},
\label{eq:pole}
\eea
for $n_l$ light flavours where $\mu$ is the arbitrary subtraction point and $a_s\equiv \alpha_s / \pi$
\section{Tests of the Factorization Assumption}
\subsection*{$\bullet$ Factorization test for PT$\oplus$NP contributions at LO }
\begin{figure}[hbt] 
\begin{center}
{\includegraphics[width=5cm  ]{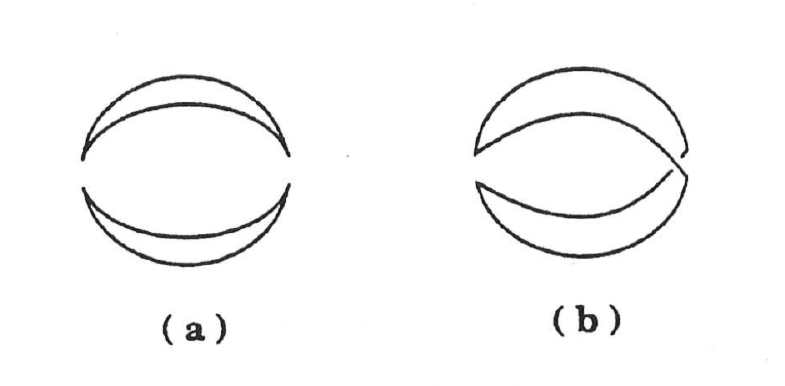}}
\caption{
\scriptsize 
{\bf (a)} Factorized contribution to the four-quark correlator at lowest order of PT; {\bf (b)} Non-factorized contribution at lowest order of PT (the figure comes from\, \cite{PICH}).
}
\label{fig:factor} 
\end{center}
\end{figure} 
From our previous work\,\cite{SNX2,SNCHI2}, we have noticed that assuming a factorization of the PT at LO and including NP contributions induces an effect about $2.2 \%$ for the decay constant and $0.5 \%$ for the mass, which is quite tiny. However, to avoid this (small) effect, we shall work in the following with the full non-factorized PT$\oplus$NP of the LO expressions.
\subsection*{$\bullet$ Test at NLO of PT from the $\bar{B}^0B^0$ four-quark correlator }
\begin{figure}[hbt] 
\begin{center}
{\includegraphics[width=6cm  ]{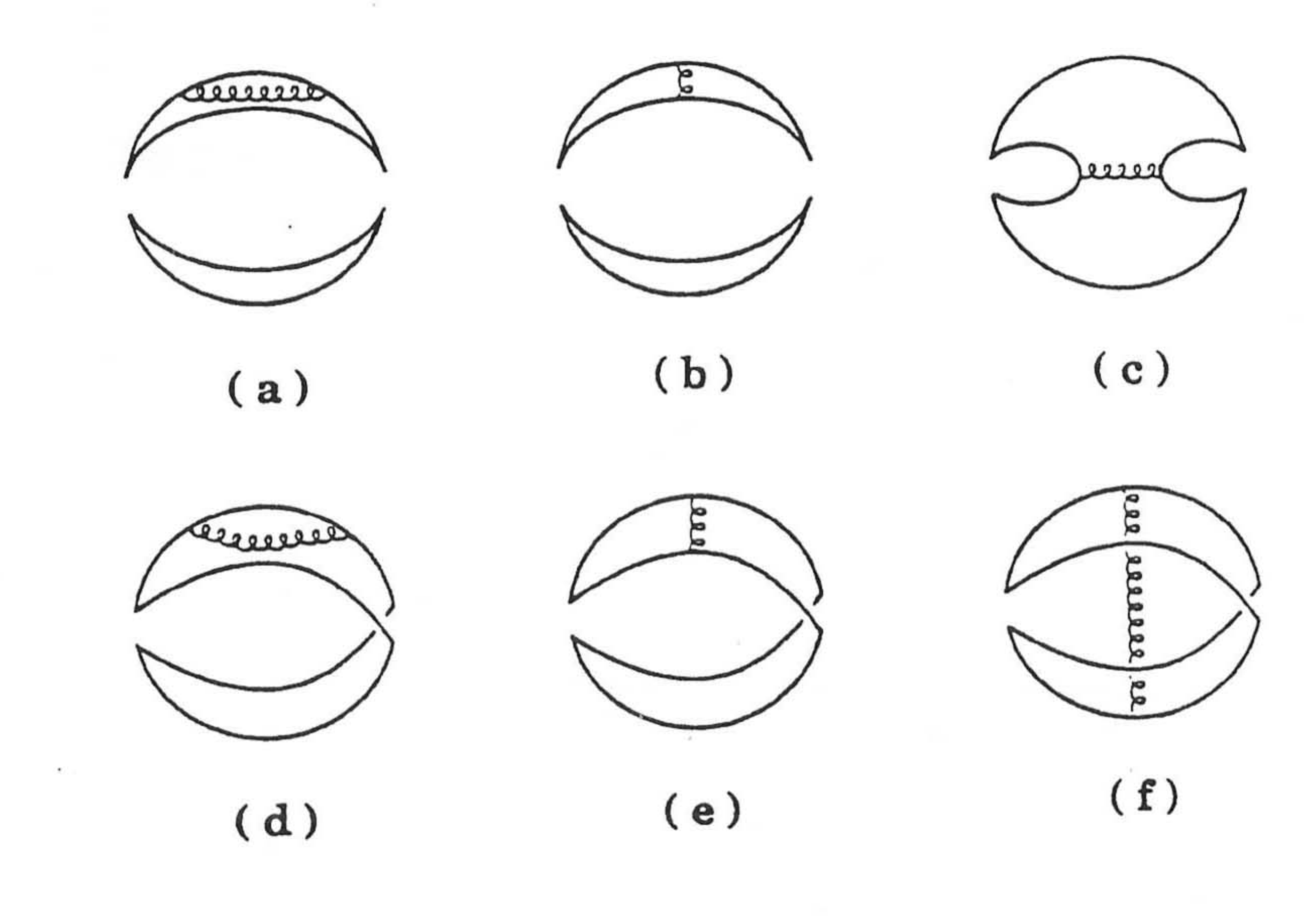}}
\caption{
\scriptsize 
{\bf (a,b)} Factorized contribution to the four-quark correlator at NLO of PT; {\bf (c to f)} Non-factorized contribution at NLO of PT (the figure comes from\, \cite{PICH}).
}
\label{fig:factoras} 
\end{center}
\end{figure} 
For extracting the PT $\alpha^n_s$ corrections to the correlator and due to the technical complexity of the calculations, we shall assume that these radiative corrections are dominated by the ones from factorized diagrams while we neglect the ones from non-factorized diagrams. This fact has been proven explicitly by\, \cite{BBAR2, BBAR3} in the case of $\bar{B}^0B^0$ systems (very similar correlator as the ones discussed in the following) where the non-factorized $\alpha_s$ corrections do not exceed $10 \%$ of the total $\alpha_s$ contributions
\subsection*{$\bullet$ Conclusions of the factorization tests}
We expect from the previous LO examples that the masses of the molecules are known with a good accuracy while, for the coupling, we shall have in mind the systematics induced by the radiative corrections estimated by keeping only the factorized diagrams. The contributions of the factorized diagrams will be extracted from the convolution integrals given in Eq. 5. Here, the suppression of the NLO corrections will be more pronounced for the extraction of the meson masses from the ratio of sum rules compared to the case of the $\bar{B}^0B^0$ systems.
\section{The $0^{++}$ and $1^{+\pm}$ Molecule States}
We shall study the charm channel and their beauty analogue. Noticing that the qualitative behaviours of the curves in these channels are very similar, we shall illustrate the analysis in the case of $\bar D_s D_s$
\subsection*{$\bullet$ $\tau$ and $t_c$ stabilities}
\begin{figure}[hbt] 
\begin{center}
{\includegraphics[width=3.8cm  ]{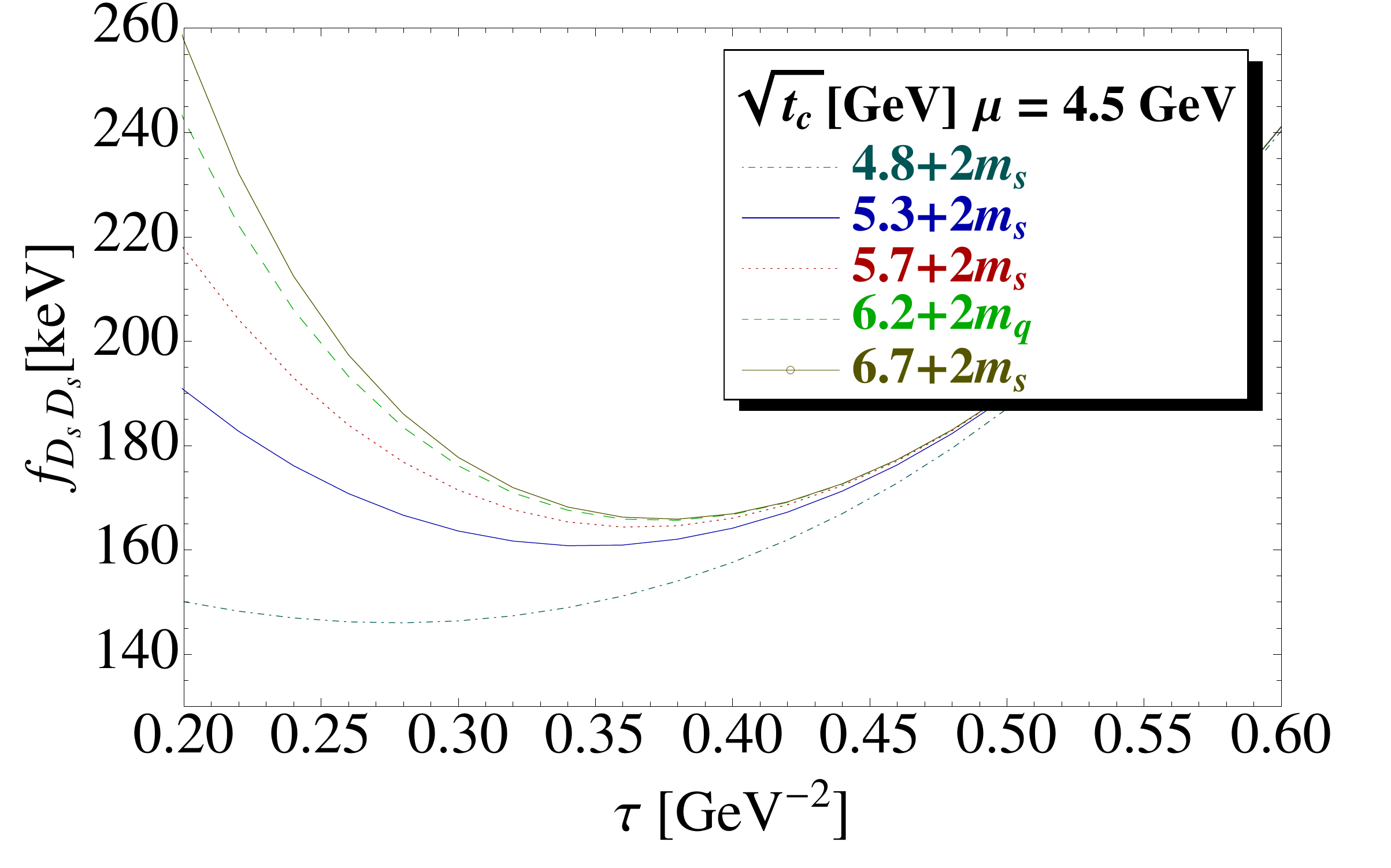}}
{\includegraphics[width=3.8cm  ]{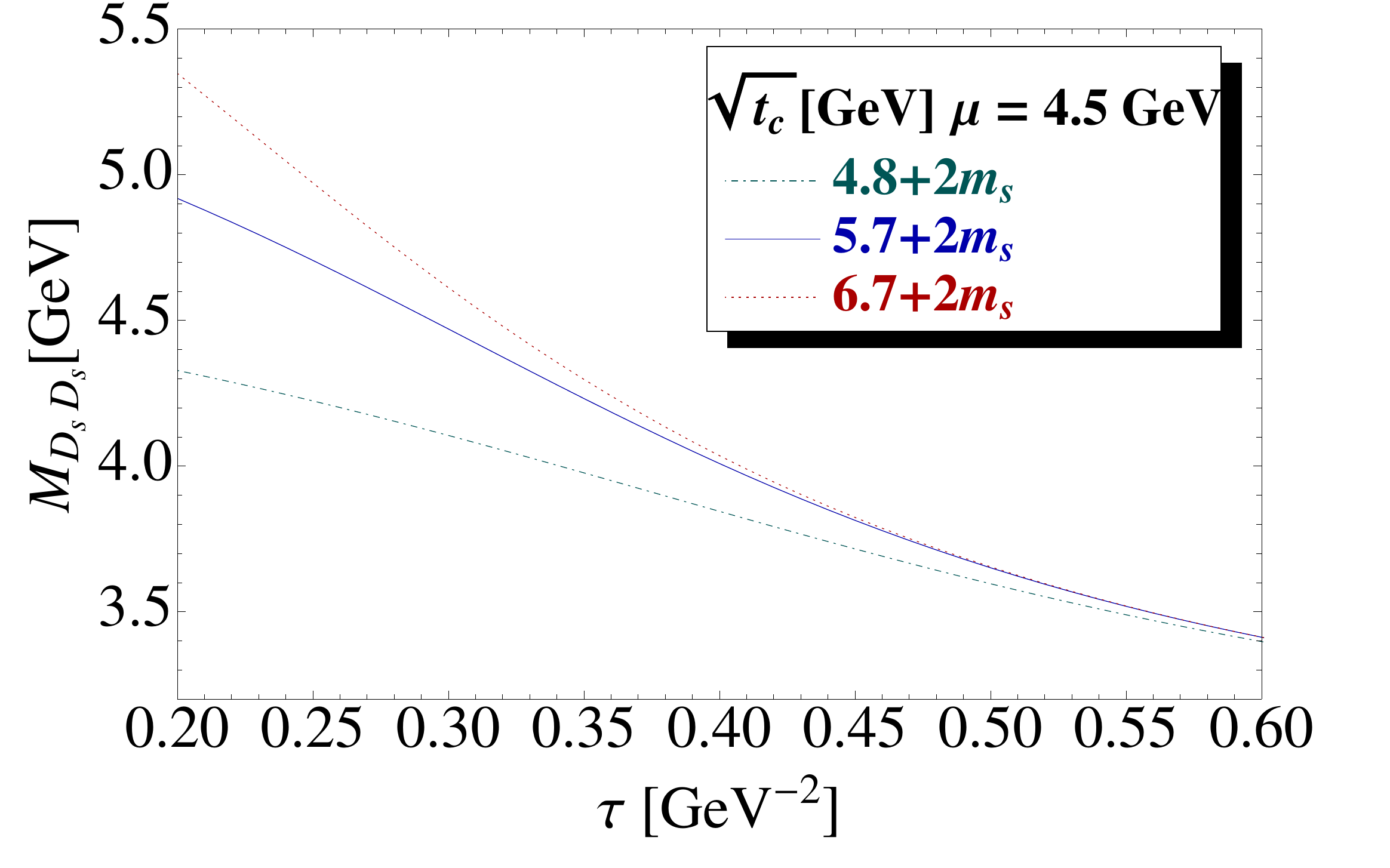}}
\centerline {\hspace*{-0.5cm} a)\hspace*{3cm} b) }
\caption{
\scriptsize 
{\bf a)} The coupling $f_{D_s D_s}$  at NLO as function of $\tau$ for different values of $t_c$, for $\mu=4.5$ GeV  and for the QCD parameters in Tables\,\ref{tab:param}; {\bf b)} The same as a) but for the mass $M_{D_s D_s}$.
}
\label{fig:fm-nlo} 
\end{center}
\end{figure}

\begin{figure}[H] 
\begin{center}
{\includegraphics[width=3.8cm  ]{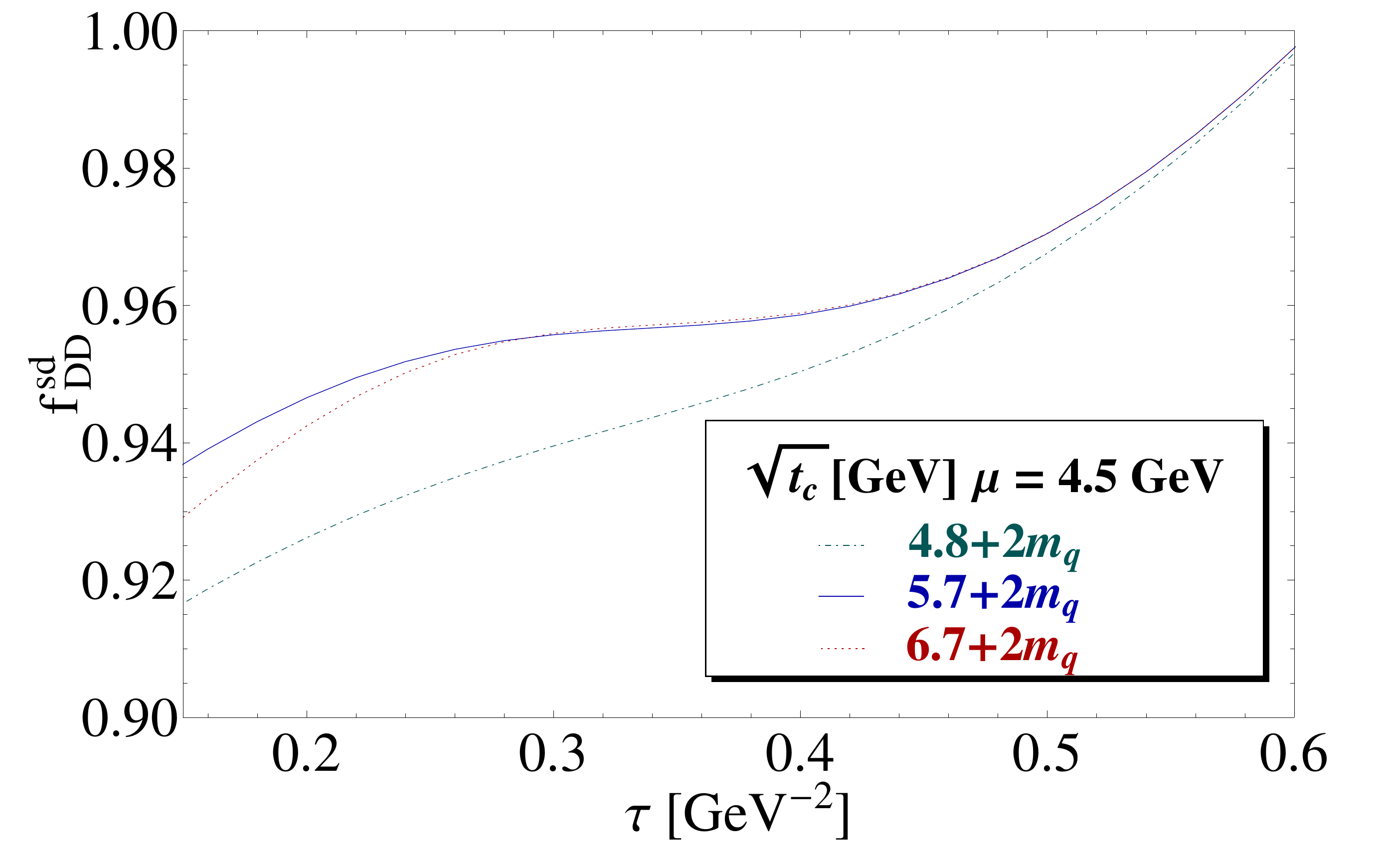}}
{\includegraphics[width=3.8cm  ]{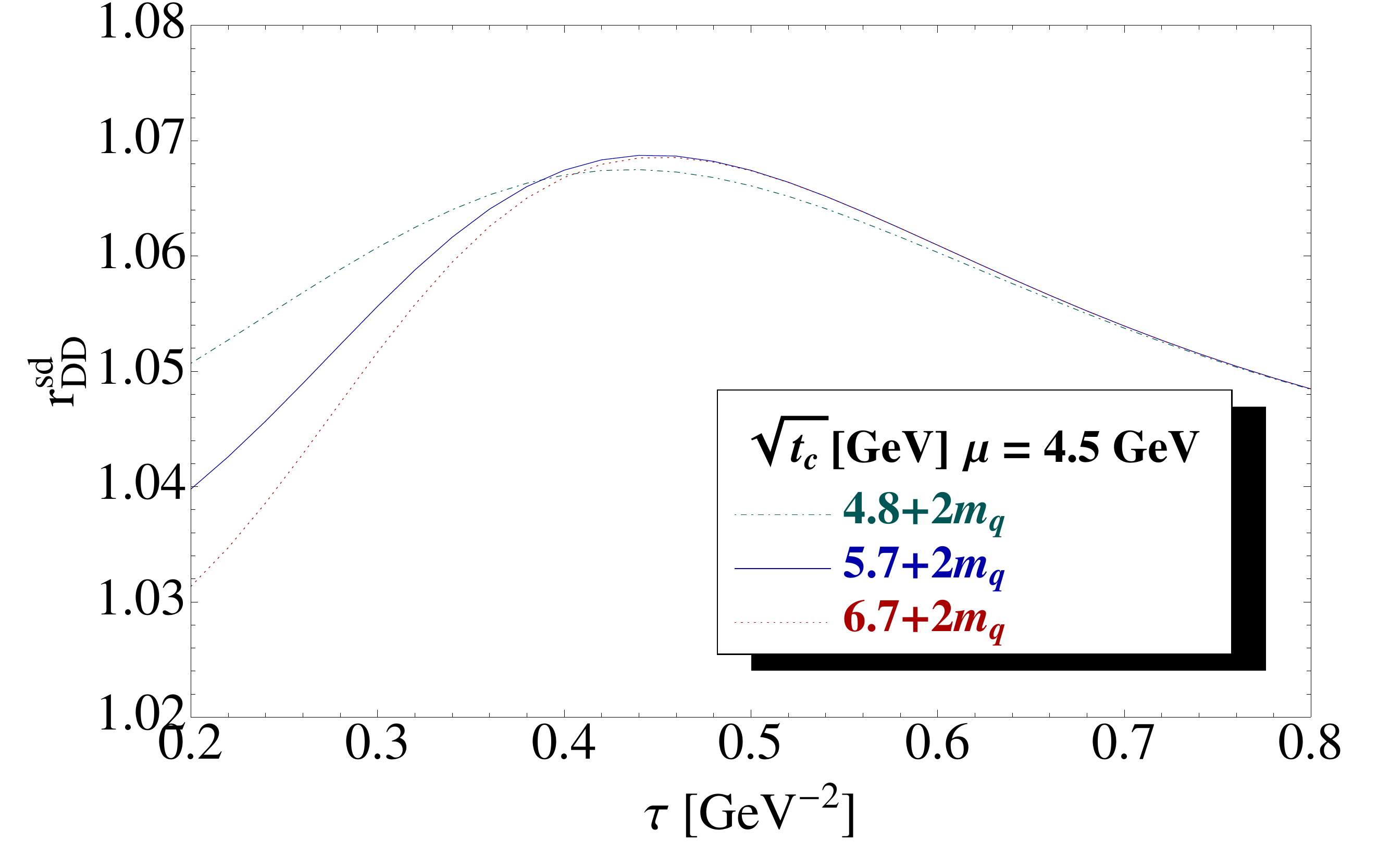}}
\centerline {\hspace*{-0.5cm} a)\hspace*{3cm} b) }
\caption{
\scriptsize 
{\bf a)} The SU3 ratio of couplings $f^{sd}_{DD}$ at NLO  as function of $\tau$ for different values of $t_c$, for $\mu=4.5$ GeV  and for the QCD parameters in Tables\,\ref{tab:param}; {\bf b)} The same as a) but for the ratio of mass $r^{sd}_{DD}$.
}
\label{fig:fmsd-nlo} 
\end{center}
\end{figure}

We study the behaviour of the coupling\footnote{Here and in the following: decay constant is the same as coupling} $f_{\bar D_sD_s}$ (resp. mass $M_{\bar D_sD_s}$) and their SU3 ratios $f^{sd}_{DD}$ (resp. $r^{sd}_{DD}$) in terms of LSR variable $\tau$ at different values of $t_c$ at NLO as shown in Fig.\ref{fig:fm-nlo} and  Fig.\ref{fig:fmsd-nlo}. We consider as an optimal estimate the mean value of coupling, mass and their SU3 ratios obtained at the minimum or inflexion point for the common range of $t_c$-values ($\sqrt{t_c}\simeq 4.8+2\overline{m}_s$ GeV) correspondig to the starting of the $\tau$-stability ($f^{sd}_{DD}$) and the one where (almost) $t_c$-stability ($\sqrt{t_c}\simeq 6.7+2\overline{m}_s$ GeV) is reached for $\tau\simeq (0.38\pm 0.02)\, \mbox{GeV}^{-2}$. In this stability regions, the requirement that the pole contribution is larger than the one of the continuum is automatically satisfied.
\subsection*{$\bullet$ $\mu$ stability}
The analysis of the $\mu$ subtraction point behaviour of the $\bar D_s D_s$ coupling and mass is very similar to the chiral limit case discussed in detail in\, \cite{SNX2}. We use the optimal choice obtained there: $\mu=(4.5\pm0.5)\, \mbox{GeV}$.
\nin
 
\section{The $0^{-\pm}$ and $1^{-\pm}$ Molecule States}
The behaviours of curves in these channels are very similar, we shall illustrate the analysis in the cases of $\bar D_{s0}^*D_s$ and  $\bar{D}_{s}^*D_{s1}$ .
\subsection*{  \b $\bar D_{s0}^*D_s$ molecule state}
\begin{figure}
\begin{center}
\includegraphics[width=3.9cm]{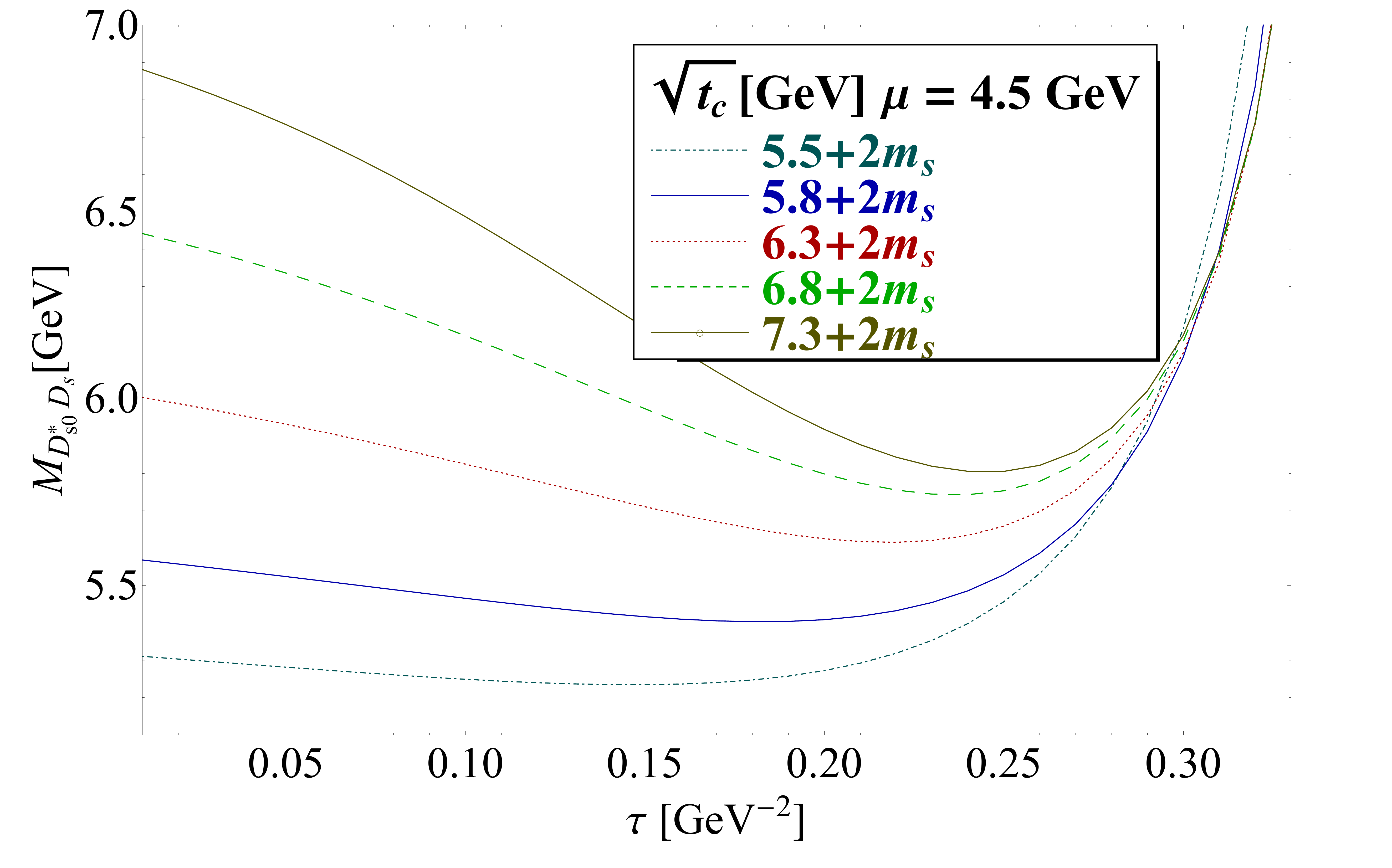}
\includegraphics[width=3.8cm]{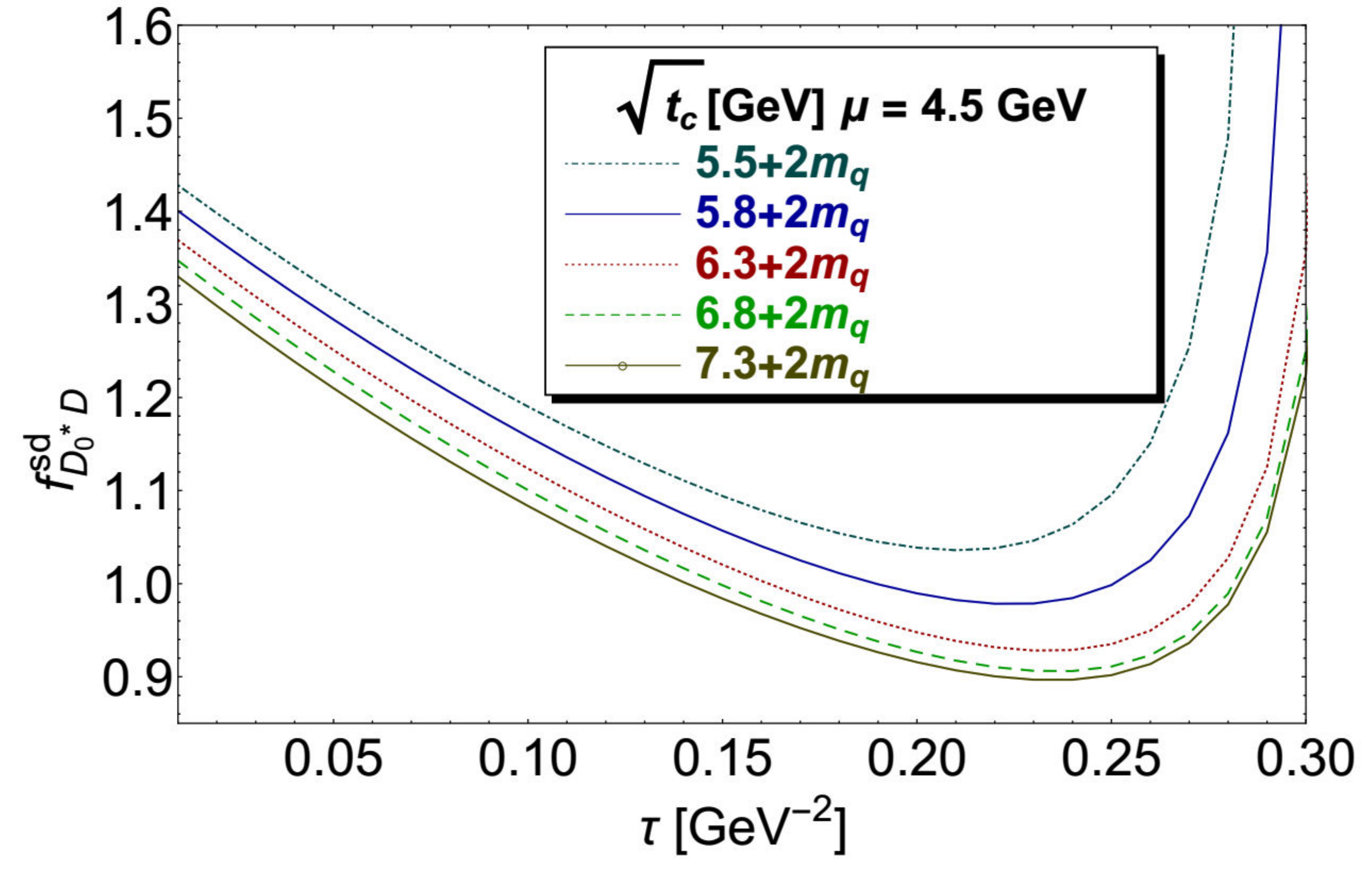}
\centerline {\hspace*{-0.5cm} a)\hspace*{3cm} b) }
\caption{\scriptsize a) The mass $M_{\bar{D}^*_{s0}D_s}$  at NLO as function of $\tau$ for different values of $t_c$, for $\mu$=4.5 GeV; b)The same as a) but for the SU3 ratio of couplings  $f^{sd}_{\bar{D}^*_{0}D}$}
\label{1A} 
\end{center}
\end{figure}

Using the optimal choice of $\mu=4.5 \pm 0.5$ GeV obtained in \cite{SNX2}, the mass and SU3 ratios of couplings present minima for $\tau$=0.18 (resp.0.25) ${\rm GeV}^{-2}$ and  $\tau$=0.22 (resp.0.24) ${\rm GeV}^{-2}$, as shown in Fig.{\ref{1A}}, within the range of $t_c$  corresponding to the beginning of the $\tau$-stability for $\sqrt{t_c}=5.8 + 2 \bar m_q\,{\rm GeV}$  (resp. $ \sqrt{t_c}=7.3 + 2 \bar m_q\,{\rm GeV}$) the one where $t_c$ stability starts to be reached.
We deduce from these regions:
\bea
M_{\bar{D}^*_{s0}D_s} &\simeq& 5604(201)_{t_c}(17)_{\tau} ...{\rm MeV},\nnb \\
  f^{sd}_{\bar{D}^*_0D} &\simeq& 0.938(41)_{t_c}(2)_{\tau} ...,
\eea
using the values of coupling and mass  $f_{\bar{D}^*_{0}D}$=240(16) keV and  $M_{\bar{D}^*_{0}D}$=5800(115) MeV, from chiral limit \cite{SNX2}, we get at NLO:
\bea
 r^{sd}_{\bar{D}^*_0D}&\simeq& 0.97(2)_M(5)_{t_c}(0)_{\tau}... ,\nnb \\ 
f_{\bar{D}^*_{s0}D_s} &\simeq& 225(15)_f(10)_{t_c}(1)_{\tau}...{\rm keV},
\eea
\subsection*{ \b $\bar{D}_{s}^*D_{s1}$ molecule state}
\begin{figure}
\begin{center}
\includegraphics[width=5cm]{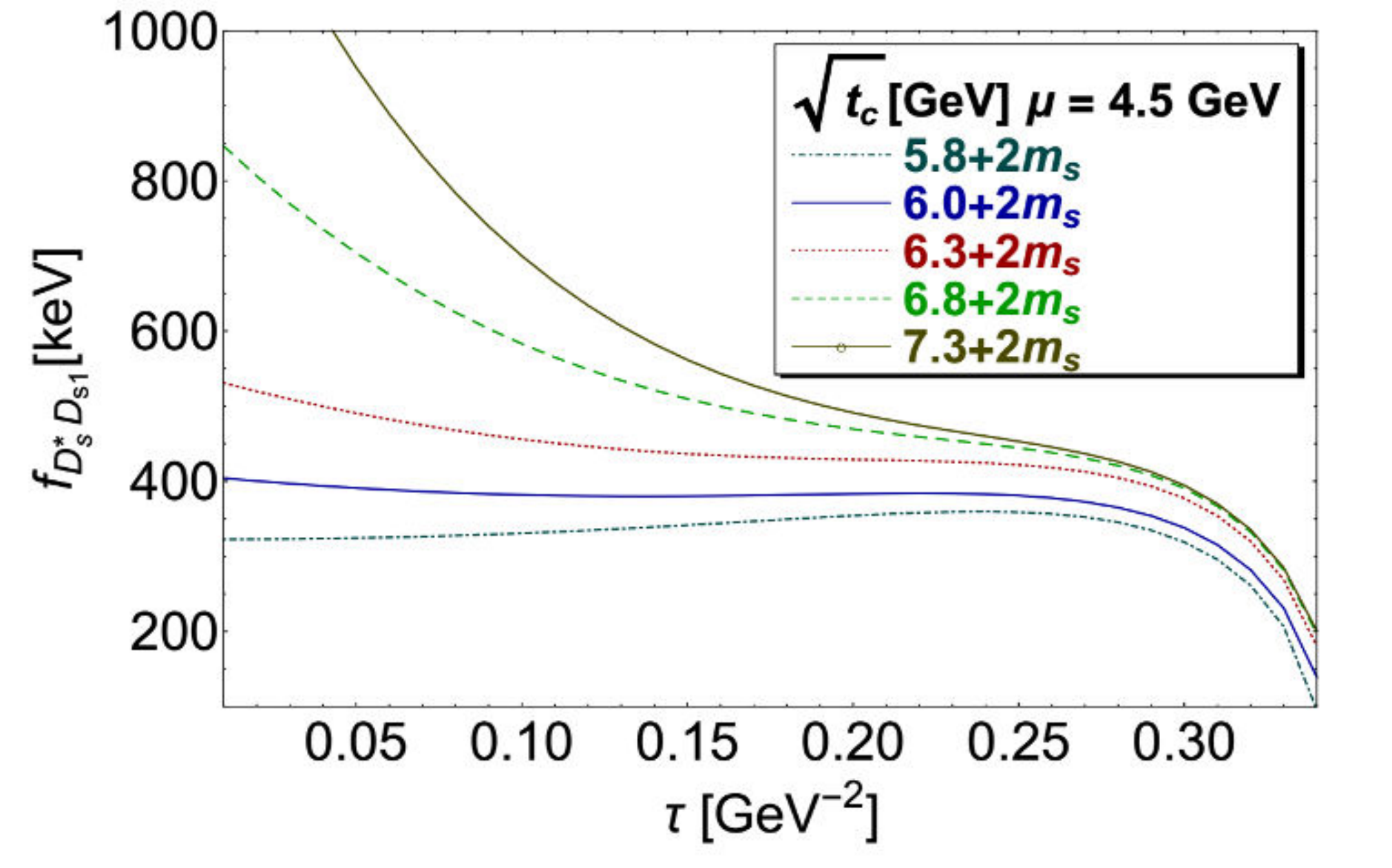}
\caption{\scriptsize $f_{\bar{D}^*_{s}D_{s1}}$  at NLO as function of $\tau$ for different values of $t_c$, for $\mu$=4.5 GeV }
\label{1D} 
\end{center}
\end{figure}

The shapes of different curves for the mass and SU3 ratio of couplings are very similar to the case of  $\bar{D}^*_{s0}D_s$ and we shall not show them here. But for this case, the coupling also presents $\tau$-stabilities as shown in Fig.\ref{1D}  from $\sqrt{t_c}=6.0+ 2 \bar{m_q}\,{\rm GeV}$ (resp. $\sqrt{t_c}=7.3+ 2 \bar{m_q}\, {\rm GeV} $) and for $\tau$=0.14 (resp. 0.21) ${\rm GeV}^{-2}$. Within the same range of $t_c$, the ratio of couplings presents stability at $\tau$=0.22 (resp. 0.24) ${\rm GeV}^{-2}$ while the minima for the mass occur at $\tau$=0.19 (resp. 0.25) ${\rm GeV}^{-2}$.\\
With the optimal results deduced from these regions and using the values  of coupling and mass $f_{D^*D_{1}}$=490(25) {\rm keV} and  $M_{D^*D_{1}}$=5898(89) MeV, from chiral limit \cite{SNX2}, we get at NLO:
\bea
 M_{D_{s}^*D_{s1}} &\simeq &5724(176)_{t_c}(14)_{\tau}... {\rm MeV}, \nnb\\
 r^{sd}_{D^*D_{1}} &\simeq& 0.97 (1.5)_M(5)_{t_c}(0)_{ \tau} ..., \nnb \\
 f_{D_{s}^*D_{s1}}&\simeq&455(22)... {\rm keV}, \nnb \\
 f^{sd}_{D^*D_{1}} &\simeq& 0.93 (1)... .
 \eea

\section{The Molecule States masses and couplings}
The results are given in Table \ref{tab:resultc} (resp. Table \ref{tab:resultb}) for the charm (resp. bottom) channel. The errors come from the QCD parameters and from the range of $\tau, t_c$ and $\mu$ where the optimal results are extracted.  We find that the SU3 breakings are relatively small for the masses ($\leq 10$ (resp. 3)  $\%$) for the charm (resp. bottom) channels while they are large ($\leq 20\, \%$) for the couplings. Like in the chiral limit case, the couplings decrease faster: $1/m_{b}^{3/2}$ than $1/m_{b}^{1/2}$ of HQET. 

\section{Four-quark states masses and couplings}
The behaviours of the corresponding curves are very similar to the previous molecule ones. The results are given in Table \ref{tab:4q-resultc} (resp. Table \ref{tab:4q-resultb}) for the charm (resp. bottom) channel. The sources of errors are the same as in the molecules case. Our conclusion is similar to the previous case of molecule states.
\section{Confrontation with some LO results and data}
\subsection*{$\bullet$ Comparison with some previous LO QSSR results}
\nin
The comparison is only informative as it is known that the LO results suffer from the ill-defined definition of the quark mass used in the analysis at this order. Most of the authors (see e.g \, \cite{CHINESE2,RABN,WLZ,QT,CHINESE3}) use the running mass value which is not justified when one implicitly uses the QCD expression obtained within the on-shell scheme. The difference between some results is also due to the way for extracting the optimal information from the analysis. Here we use well-defined stability criteria verified from the example of the harmonic oscillator in quantum mechanics and from different well-known hadronic channels.
\subsection*{$\bullet$ Confrontation with experiments}
We conclude from the previous analysis that:\\
-- The $0^{++}$ X(4700) experimental candidate might be identified with a $\bar D^*_{s0}D^*_{s0}$ molecule ground state.\\
-- The interpretation of the $0^{++}$ candidates as pure four-quark ground states is not favoured by our result.\\
-- The $1^{++}$ X(4147) and X(4273) are compatible within the error with the one of the $\bar D^*_sD_s$ molecule state and with the one of the axial-vector $A_c$ four-quark state.\\
-- Our predictions suggest the presence of $0^{++}$ $\bar D_sD_s$ and $\bar D^*_sD^*_s$ molecule states in the range $(4121\sim4396)$ MeV and a $\bar D^*_{s0}D_{s1}$ state around 4841 MeV.\\
-- We also present new predictions for the $0^{-\pm}$, $1^{-\pm}$ and for different beauty states which can be tested in future experiments. 
\section{Conclusion}
We have summarized  our results for SU3 breaking at NLO and N2LO of PT\,\cite{SU3}
for  molecule and four-quark states (see Table\,\ref{tab:resultc} to Table\,\ref{tab:4q-resultb}). They are important for further building of an effective theory for these exotic states and can be tested by lattice calculations. We plan to extend this analysis for the estimate of the meson widths.
\section*{Acknowledgements}
We thank A. Rabemananjara for participating at the early stage of this work.
{\scriptsize
\begin{table*}[H]
\setlength{\tabcolsep}{1.13pc}
 \caption{$\bar{D}D$-like molecules couplings, masses and their corresponding SU3 ratios  from LSR within stability criteria at NLO to N2LO of PT. We include revised estimates of the $ \bar D^{*}_{0}D^{*}_{0}$,  $ \bar D^{*}_{0}D_{1}$ couplings and masses and new one for $ \bar D_{1}D_{1}$.}  
{\scriptsize
\begin{tabular*}{\textwidth}{@{}ll   ll  ll  ll l@{\extracolsep{\fill}}l}
\hline
\hline
                \bf Channels &\multicolumn{2}{c}{$f^{sd}_M\equiv f_{M_s}/f_{M}$}
					&\multicolumn{2}{c}{$f_{M_s}$\bf [keV]}
					&\multicolumn{2}{c}{$r^{sd}_M\equiv M_{M_s}/M_{M}$}
					&\multicolumn{2}{c}{$M_{M_s}$ \bf [MeV]}\\
\cline{2-3} \cline{4-5}\cline{6-7}\cline{8-9}
                 & \multicolumn{1}{l}{{NLO}}
                 & \multicolumn{1}{l }{N2LO} 
                 & \multicolumn{1}{l}{NLO} 
                 & \multicolumn{1}{l }{N2LO} 
                 & \multicolumn{1}{l}{NLO} 
                 & \multicolumn{1}{l}{N2LO}
		    & \multicolumn{1}{l}{NLO} 
                 & \multicolumn{1}{l}{N2LO} 
                  \\
\hline
 \bf{Scalar($0^{++}$)}&&&&&&&&\\
$\bar D_sD_s$&$0.95(3)$&0.98(4)&156(17)&167(18)&1.069(4)&1.070(4)&4169(48)&4169(48)\\
$\bar D^{*}_{s}D^{*}_{s}$&0.93(3)&0.95(3)&265(31)&284(34)&1.069(3)&1.075(3)&4192(200)&4196(200)\\
$\bar D^{*}_{s0}D^{*}_{s0}$&0.88(6)&0.89(6)&85(12)&102(14)&1.069(69)&1.058(68)&4277(134)&4225(132)\\
$\bar D_{s1}D_{s1}$&0.906(33)&0.930(34)&209(28)&229(31)&1.097(7)&1.090(7)&4187(62)&4124(61) \\
\\
$\bar D^{*}_{0}D^{*}_{0}$&--&--&97(15)&114(18)&--&--&4003(227)&3954(224)\\
$\bar D_{1}D_{1}$&--&--&236(32)&274(37)&--&--&3838(57)&3784(56) \\
\bf {Axial($1^{+\pm}$)}&&&&&&&&\\
$\bar D^{*}_{s}D_{s}$&0.93(3)&0.97(3)&143(16)&156(17)&1.070(4)&1.073(4)&4174(67)&4188(67)\\
$\bar D^{*}_{s0}D_{s1}$&0.90(1)&0.82(1)&87(14)&110(18)&1.119(24)&1.100(24)&4269(205)&4275(206)\\
\\
$\bar D^{*}_{0}D_{1}$&--&--&96(15)&112(17)&--&--&3849(182)&3854(182)\\
\bf  {Pseudo($0^{-\pm}$)}&&&&&&&&\\
$\bar D^{*}_{s0}D_{s}$&0.94(5)&0.90(4)&225(24)&232(25)&0.970(50)&0.946(40)&5604(223)&5385(214)\\
$\bar D^{*}_{s}D_{s1}$&0.93(4)&0.90(4)&455(34)&508(38)&0.970(50)&0.972(34)&5724(195)&5632(192)\\
\bf {Vector($1^{--}$)} &&&&&&&&\\
$\bar D^{*}_{s0}D^{*}_{s}$&0.87(4)&0.86(4)&208(11)&216(11)&0.980(33)&0.956(32)&5708(184)&5571(180)\\
$\bar D_{s}D_{s1}$&0.97(3)&0.93(3)&202(12)&213(13)&0.970(33)&0.951(31)&5459(122)&5272(120)\\
\bf { Vector($1^{-+}$)} &&&&&&&&\\
$\bar D^{*}_{s0}D^{*}_{s}$&0.98(5)&0.92(5)&219(17)&231(18)&0.963(32)&0.948(32)&5699(184)&5528(179)\\
$\bar D_{s}D_{s1}$&0.92(3)&0.88(3)&195(13)&212(14)&0.959(34)&0.955(34)&5599(155)&5487(152)\\
\end{tabular*}
}
\label{tab:resultc}
\end{table*}
}
\vspace*{-0.3cm}
{\scriptsize
\begin{table*}[H]
\setlength{\tabcolsep}{1.08pc}
 \caption{$\bar{B}B$-like molecules couplings, masses and their corresponding SU3 ratios  from LSR within stability criteria at NLO to N2LO of PT.  The * indicates that the value does not come from a direct determination.
}
{\scriptsize
\begin{tabular*}{\textwidth}{@{}ll   ll  ll  ll l@{\extracolsep{\fill}}l}
\hline
\hline
                \bf Channels &\multicolumn{2}{c}{$f^{sd}_M\equiv f_{M_s}/f_{M}$}
					&\multicolumn{2}{c}{$f_{M_s}$\bf [keV]}
					&\multicolumn{2}{c}{$r^{sd}_M\equiv M_{M_s}/M_{M}$}
					&\multicolumn{2}{c}{$M_{M_s}$ \bf [MeV]}\\
\cline{2-3} \cline{4-5}\cline{6-7}\cline{8-9}
                 & \multicolumn{1}{l}{NLO} 
                 & \multicolumn{1}{l }{N2LO} 
                 & \multicolumn{1}{l}{NLO} 
                 & \multicolumn{1}{l }{N2LO} 
                 & \multicolumn{1}{l}{NLO} 
                 & \multicolumn{1}{l}{N2LO}
		    & \multicolumn{1}{l}{NLO} 
                 & \multicolumn{1}{l}{N2LO} 
                  \\
\hline
 \bf{Scalar($0^{++}$)}&&&&&&&&\\
$\bar B_sB_s$&1.04(4)&1.15(4)&17(2)&20(2)&1.027(4)&1.029(4)&10884(74)&10906(74)\\
$\bar B^{*}_{s}B^{*}_{s}$&1.00(3)&1.12(3)&31(5)&36(6)&1.028(5)&1.029(5)&10944(134)&10956(134)\\
$\bar B^{*}_{s0}B^{*}_{s0}$&1.11(5)&1.07(5)&13(3)&17(4)&1.050(11)&1.034(11)&11182(227)&11014(224)\\
$\bar B_{s1}B_{s1}$&1.197(73)&1.214(74)&24(5)&29(6)&1.040(2)&1.035(2)&10935(170)&10882(169) \\
 \\
$\bar B_{1}B_{1}$&--&--&20(3)&28.6(4)&--&--&10514(149)&10514(149) \\
\bf {Axial($1^{+\pm}$)}&&&&&&&&\\
$\bar B^{*}_{s}B_{s}$&1.01(3)&1.18(4)&16.7(2)&20(2)&1.028(4)&1.030(4)&10972(195)&10972(195)\\
$\bar B^{*}_{s0}B_{s1}$&0.80(4)&0.79(4)&9.1(2.2)&10.7(2.6)&1.052(14)&1.031(14)&11234(208)&11021(204)\\
\bf  {Pseudo($0^{-\pm}$)}&&&&&&&&\\
$\bar B^{*}_{s0}B_{s}$&1.06(3)&1.02(3)&58(3)&68(4)&1.00(3)*&1.00(3)*&12725(217)&12509(213)\\
$\bar B^{*}_{s}B_{s1}$&0.96(4)&0.95(4)&100(11)&118(13)&1.00(3)*&1.00(3)*&12726(295)&12573(292)\\
\bf { Vector($1^{--}$)} &&&&&&&&\\
$\bar B^{*}_{s0}B^{*}_{s}$&0.95(3)&0.90(3)&51(4)&59(5)&1.00(3)*&0.99(3)*&12715(267)&12512(263)\\
$\bar B_{s}B_{s1}$&0.83(4)&0.77(3)&45(3)&50(3)&0.99(3)*&0.99(3)*&12615(236)&12426(233)\\
\bf { Vector($1^{-+}$)} &&&&&&&&\\
$\bar B^{*}_{s0}B^{*}_{s}$&0.94(3)&0.92(3)&51(5)&59(6)&1.00(3)*&0.99(3)*&12734(262)&12479(257)\\
$\bar B_{s}B_{s1}$&0.89(4)&0.85(3)&48(5)&55(6)&0.99(3)*&0.98(3)*&12602(247)&12350(242)\\
\end{tabular*}
}
\label{tab:resultb}
\end{table*}
}
\vspace*{-0.3cm}
{\scriptsize
\begin{table*}[H]
\setlength{\tabcolsep}{1.2pc}
 \caption{4-quark couplings, masses and their corresponding SU3 ratios  from LSR within stability criteria at NLO and N2LO of PT. The * indicates that the value does not come from a direct determination. 
}  
{\scriptsize{
\begin{tabular*}{\textwidth}{@{}ll   ll  ll  ll l@{\extracolsep{\fill}}l}
\hline
\hline
                \bf Channels &\multicolumn{2}{c}{$f^{sd}_M\equiv f_{M_s}/f_{M}$}
					&\multicolumn{2}{c}{$f_{M_s}$\bf [keV]}
					&\multicolumn{2}{c}{$r^{sd}_M\equiv M_{M_s}/M_{M}$}
					&\multicolumn{2}{c}{$M_{M_s}$ \bf [MeV]}\\
\cline{2-3} \cline{4-5}\cline{6-7}\cline{8-9}
                 & \multicolumn{1}{l}{{NLO}}
                 & \multicolumn{1}{l }{N2LO} 
                 & \multicolumn{1}{l}{NLO} 
                 & \multicolumn{1}{l }{N2LO} 
                 & \multicolumn{1}{l}{NLO} 
                 & \multicolumn{1}{l}{N2LO}
		    & \multicolumn{1}{l}{NLO} 
                 & \multicolumn{1}{l}{N2LO} 
                  \\
\hline
 \bf{c-quark}&&&&&&&&\\
$S_{sc}(0^+)$&0.91(4)&0.98(4)&161(17)&187(19)&1.085(11)&1.086(11)&4233(61)&4233(61)\\
$A_{sc}(1^+)$&0.80(4)&0.87(4)&141(15)&160(17)&1.081(4)&1.082(4)&4205(112)&4209(112)\\
$\pi_{sc}(0^-)$&0.88(7)&0.86(7)&256(29)&267(30)&0.97(3)*&0.96(3)*&5671(181)&5524(176)\\
$V_{sc}(1^-)$&0.91(10)&0.87(10)&245(31)&258(33)&0.96(4)*&0.96(4)*&5654(239)&5539(234)\\
\hline
\hline
\end{tabular*}
}}
\label{tab:4q-resultc}
\end{table*}
}
\vspace{-0.3cm}
{\scriptsize
\begin{table*}[H]
\setlength{\tabcolsep}{1.22pc}
 \caption{4-quark couplings, masses and their corresponding SU3 ratios  from LSR within stability criteria at NLO and N2LO of PT. The * indicates that the value does not come from a direct determination. 
}  
{\scriptsize{
\begin{tabular*}{\textwidth}{@{}ll   ll  ll  ll l@{\extracolsep{\fill}}l}
\hline
\hline
                \bf Channels &\multicolumn{2}{c}{$f^{sd}_M\equiv f_{M_s}/f_{M}$}
					&\multicolumn{2}{c}{$f_{M_s}$\bf [keV]}
					&\multicolumn{2}{c}{$r^{sd}_M\equiv M_{M_s}/M_{M}$}
					&\multicolumn{2}{c}{$M_{M_s}$ \bf [MeV]}\\
\cline{2-3} \cline{4-5}\cline{6-7}\cline{8-9}
                 & \multicolumn{1}{l}{{NLO}}
                 & \multicolumn{1}{l }{N2LO} 
                 & \multicolumn{1}{l}{NLO} 
                 & \multicolumn{1}{l }{N2LO} 
                 & \multicolumn{1}{l}{NLO} 
                 & \multicolumn{1}{l}{N2LO}
		    & \multicolumn{1}{l}{NLO} 
                 & \multicolumn{1}{l}{N2LO} 
                  \\
\hline
 \bf{b-quark}&&&&&&&&\\
$S_{sb}(0^+)$&0.78(3)&0.83(3)&22(5)&26(6)&1.044(4)&1.048(4)&11122(149)&11133((149)\\
$A_{sb}(1^+)$&0.92(3)&0.98(3)&22(4)&26(5)&1.042(6)&1.046(6)&11150(172)&11172(172)\\
$\pi_{sb}(0^-)$&0.80(7)&0.76(4)&66(12)&71(13)&0.985(2)*&0.975(2)*&12730(215)&12374(209)\\
$V_{sb}(1^-)$&0.97(6)&0.90(6)&64(8)&68(9)&0.996(3)*&0.984(30)*&12716(272)&12411(266)\\
\hline
\hline
\end{tabular*}
}}
\label{tab:4q-resultb}
\end{table*}
}

\end{document}